\documentclass[preprint,journal]{vgtc}       % preprint (journal style)

\ifpdf%                                % if we use pdflatex
  \pdfoutput=1\relax                   % create PDFs from pdfLaTeX
  \usepackage{graphicx}                % allow us to embed graphics files
  \DeclareGraphicsExtensions{.pdf,.png,.jpg,.jpeg} % for pdflatex we expect .pdf, .png, or .jpg files
\else%                                 % else we use pure latex
  \usepackage{graphicx}                % allow us to embed graphics files
  \DeclareGraphicsExtensions{.eps}     % for pure latex we expect eps files
\fi%

\graphicspath{{figures/}{pictures/}{images/}{./}} % where to search for the images

\usepackage{microtype}                 % use micro-typography (slightly more compact, better to read)
\PassOptionsToPackage{warn}{textcomp}  % to address font issues with \textrightarrow
\usepackage{textcomp}                  % use better special symbols
\usepackage{mathptmx}                  % use matching math font
\usepackage{times}                     % we use Times as the main font
\usepackage{cite}                      % needed to automatically sort the references
\usepackage{tabu}                      % only used for the table example
\usepackage{booktabs}                  % only used for the table example
\usepackage{xcolor}
\usepackage{enumitem}
\usepackage{mathtools}
\usepackage{tabularx}
\usepackage{ wasysym }

\preprinttext{To appear in IEEE Transactions on Visualization and Computer Graphics, \url{https://doi.org/10.1109/TVCG.2020.2969007}.}

\onlineid{0}

\vgtccategory{Research}
\vgtcpapertype{system}

\title{Security in Process:\\Visually Supported Triage Analysis in Industrial Process Data}

\author{Anna-Pia Lohfink, Simon D. Duque Anton, Hans Dieter Schotten, Heike Leitte, \textit{Member, IEEE}, \\and Christoph Garth, \textit{Member, IEEE}}
\authorfooter{
\item
 Anna-Pia Lohfink is with the Scientific Visualization Lab, University of Kaiserslautern. E-mail: lohfink@cs.uni-kl.de.
\item
 Simon D. Duque Anton is with the German Research Center for Artificial Intelligence Kaiserslautern. E-mail: simon.duque\_anton@dfki.de.
\item
 Hans Dieter Schotten is with the German Research Center for Artificial Intelligence Kaiserslautern. E-mail: hans\_dieter.schotten@dfki.de.
 \item Heike Leitte is with the Visual Information Analysis Group, University of Kaiserslautern. E-mail: leitte@cs.uni-kl.de
\item Christoph Garth is with the Scientific Visualization Lab, University of Kaiserslautern. E-mail: garth@cs.uni-kl.de.
}

\shortauthortitle{Lohfink \MakeLowercase{\textit{et al.}}: Security in Process: Visually Supported Triage Analysis in Industrial Process Data}
\abstract{Operation technology networks, i.e. hard- and software used for monitoring and controlling physical/industrial processes, have been considered immune to cyber attacks for a long time. A recent increase of attacks in these networks proves this assumption wrong. Several technical constraints lead to approaches to detect attacks on industrial processes using available sensor data. This setting differs fundamentally from anomaly detection in IT-network traffic and requires new visualization approaches adapted to the common periodical behavior in OT-network data. We present a tailored visualization system that utilizes inherent features of measurements from industrial processes to full capacity to provide insight into the data and support triage analysis by laymen and experts. The novel combination of spiral plots with results from anomaly detection was implemented in an interactive system. The capabilities of our system  are demonstrated using sensor and actuator data from a real-world water treatment process with introduced attacks. Exemplary analysis strategies are presented. Finally, we evaluate effectiveness and usability of our system and perform an expert evaluation.%
} % end of abstract

\keywords{Cyber Security, Information Visualization, Anomaly Detection, Triage Analysis, Operation Technology Networks}
\CCScatlist{ % not used in journal version
 \CCScat{K.6.1}{Management of Computing and Information Systems}%
{Project and People Management}{Life Cycle};
 \CCScat{K.7.m}{The Computing Profession}{Miscellaneous}{Ethics}
}

\teaser{
  \centering
  \includegraphics[width=\linewidth]{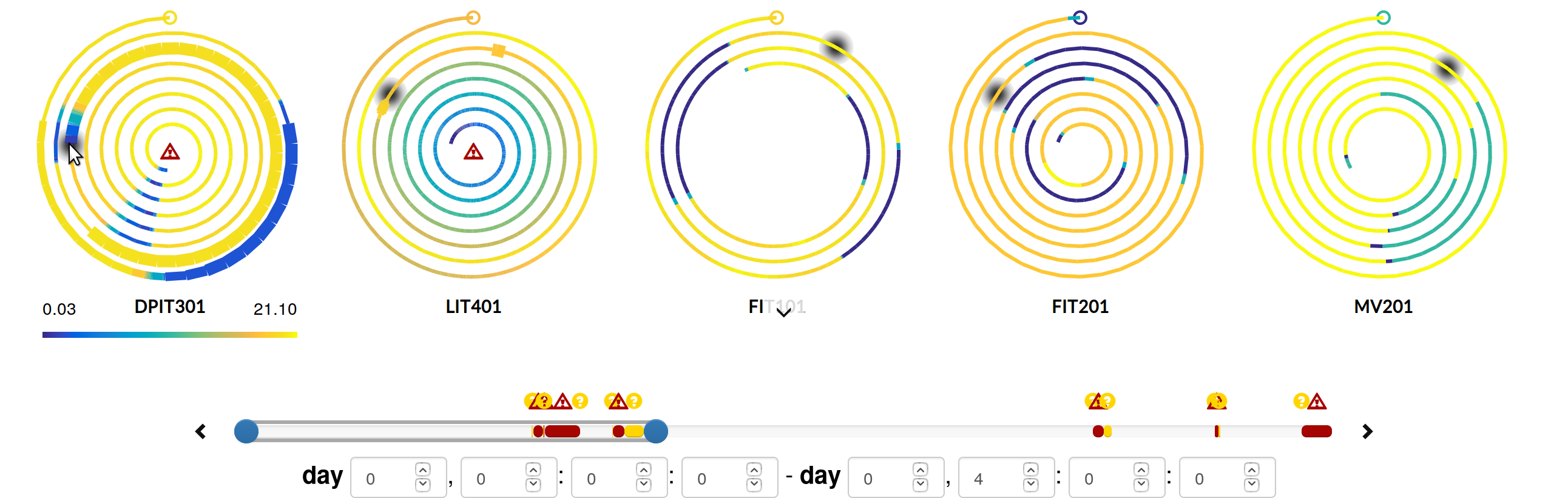}
  \caption{Overview of our visualization system: To support triage analysis in OT-networks, our visualization combines sensor readings with results from an anomaly detection algorithm. Spiral plots encode readings and detection results in color and line thickness respectively. An overview of the complete time frame is provided in a time slider and anomalies are highlighted.}
	\label{fig:teaser}
}

\vgtcinsertpkg

\begin{document}


\firstsection{Introduction}

\maketitle

\newcommand{\spiralheight}{3.8cm}

%% \section{Introduction} %for journal use above \firstsection{..} instead
Over the last two decades, automation and thus the use of Operation Technology (OT) networks in industry have increased rapidly, as have attacks on such networks  \cite{anton2017two}. Network environments that are difficult to update and the use of communication protocols that do not contain authentication or encryption \cite{Modbus:2012,Modbus:2006} lead to high vulnerability once an intruder has successfully breached the communication network. In addition, historical reasons caused these networks to be less secured against attacks than deemed appropriate for home and office IT \cite{IGURE2006498}. First, OT and IT networks are supposed to be physically separated. Second, attacking OT networks is expected to be difficult due to their highly application specific implementation. 
However, both of these aspects have become obsolete nowadays. Commercial off-the-shelf products in the industrial area such as \textit{Programmable Logic Controllers} (PLCs) facilitate set up, maintenance and operation of industrial applications by using common interfaces and programming libraries. This also decreases the difficulties for attackers. Furthermore, new use and business cases introduced as part of Industry 4.0 efforts \cite{3gpp2017} break the physical separation of networks. Relying on the communication and computation capabilities of (industrial) internet of things devices, access routes to OT networks are created. Even if no such access is possible, attackers have successfully managed to move laterally to the OT networks in the past after breaking the IT network perimeter \cite{Duque_Anton.2019c}.\\
This lack in security has lead to recent research in cyber security with the aim to detect attacks in already available information, such as sensors monitoring the production process \cite{anton2019implementing}. In network security, periodic behavior is a signal for malware activity \cite{huynh2016periodic}. In contrast, sensor data monitoring ongoing industrial processes oftentimes develop periodic patterns, such that the absence of periodicity represents an anomaly. Exploiting this fact, we developed a time-series visualization based on spiral plots that combines sensor data with results from anomaly detection. 
Thus providing not only an overview of the measurements in normal operation but also indicating possible attacks, easy to understand not only for cyber security trained personnel, but also for non-security experts. Hence, an improvement of security by supervision of the data is also possible in factories without dedicated cyber security personnel. The visualization was designed and evaluated to support triage analysis. 
\\
In particular, our contributions are:
\begin{itemize} \setlength\itemsep{-0.3em}
	\item We present a system for triage analysis that combines sensor-data visualization and anomaly detection visualization. 
	\item Analysis strategies are illustrated, that can be used to verify attacks, analyze their causes, and to detect false alarms.
	\item We apply our system to real-world data from an OT-network in an industrial water treatment process. 
	\item We evaluate effectiveness and usability of the system for security visualization experts as well as for laymen.
\end{itemize}

\begin{figure}[t!]
  \centering
  \includegraphics[width=0.9\columnwidth]{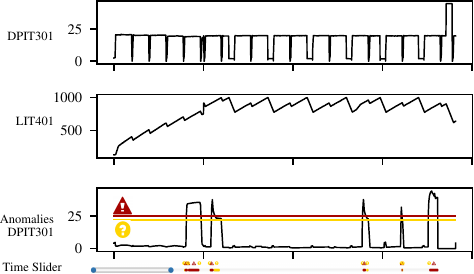}
  \caption{\label{fig:anomaly_example}Anomaly detection results: Readings of two sensors and anomaly detection results for DPIT301. Chosen thresholds for categories II and III are marked. The corresponding representation in our visualization system is given.}
\end{figure}

\section{Related Work}
Our system combines methods for time series visualization, anomaly detection, and detection and visualization of periodic behavior in monitored networks. In the following, we briefly review relevant prior work and discuss differences to our approach.
\subsection {Time Series Visualization}
Considering results from multiple sensors for a given time span, time series data needs to be visualized. A thorough summary on visualization of periodic and aperiodic time series data is given by Aigner et al. in \cite{aigner2011visualization}. For the special case of uncorrelated periodic time series, we chose the well known spiral plot by Weber et al. \cite{weber2001visualizing}. A 3D adaption of spiral plots was given by Tominski et al. \cite{1509075}. Hu et al. use spiral plots to visualize alarm floods and their patterns in complex industrial facility monitoring \cite{8264746}. To our knowledge, spiral plots for periodic time series visualization were not yet combined with anomaly detection and visualization. 

\subsection{Anomaly detection in time series}
Industrial intrusion detection based on sensor data is a widely regarded topic in cyber security research. Anomalies in the data are detected with a plethora of different approaches. Schneider et al. use autoencoders to detect anomalies in cyber-physical system networks \cite{schneider2018high}, Goh et al. and Feng et al. use neural networks for the detection \cite{Goh.2017,Feng.2017}. One class support vector machines are presented by Maglaras et al. as a machine learning algorithm to detect novel and unknown attacks \cite{Maglaras.2014}. %Anomaly detection results presented in this work have been obtained by One-Class Support Vector Machines, Isolation Forests and Matrix Profiles.

Results from different anomaly detection approaches have been visualized in different ways. 
Stoffel et al. provide a visualization combining multiple data sources that monitor a computer network \cite{Stoffel:2013:FAT:2517957.2517966}. Their visualization relies on well-known time series visualizations. Combining these visualizations and highlighting of detected anomalies facilitate shape, correlation and pattern recognition. 

Karapistoli et al. detect selective forwarding and jamming attacks in wireless sensor networks. Findings are then incorporated in graph visualization and crossed views, providing an efficient and fast overview of the network status \cite{Karapistoli:2013:SRC:2517957.2517964}.

While Boschetti et al. refrain from incorporating information on detected anomalies directly in visualization \cite{Boschetti:2011:TVQ:2016904.2016905}, they use plots, histograms, graphs and matrix visualization to provide information on network traffic. Timesteps with a potential attack can be selected from a list. The different visualizations then show the relevant time frame thus providing information on the anomaly. 

\subsection{Detection and Visualization of Periodic Behavior} 
In OT networks, periodic behavior is the norm. In contrast, such activities in monitored IT network activity are a sign of anomalies and might indicate malware activity. Application of the discrete Fourier transform allows Gove et al. to exhibit multiple period lengths simultaneously from the monitored data \cite{gove2018visualizing}. Detected features can then be grouped and sorted by cyber security relevant features. Periodic behavior is visualized over multiple period lengths by potting the time series and giving the results of the Fourier transformation. This enables experts to compare and check for coordinated periodic activity. Huynh et al. also aim on detecting periodic malware activity in their paper ``Uncovering Periodic Network Signals of Cyber Attacks''\cite{7739581}. They present a Fourier transform based periodicity detector that is able to detect even complicated periodical traces in network traffic. Clustered alerts are then visualized (amongst others) using a circular plot to compare periodic behavior of different alerts. \\
\\
Although many visualizations for anomaly and periodicity detection exist, to our knowledge none consider detection of anomalies in sensor data of industrial processes. Visualizations designed for anomaly detection in network activity are not suitable for this task since the typical periodic behavior is not taken into account or is even classified as an indicator for malware activity.
In addition, all anomaly detection systems we have considered above aim on providing information for cyber security experts. Oftentimes an assessment of the situation by laymen is impossible due to the complexity of the visualizations and systems. Our visualization combines visualization of sensor data and anomaly detection in a way that renders monitoring and basic triage analysis by laymen possible. As such, it is suitable to monitor not only anomalies in the process but also the normal behavior. Operators supervising the considered process are able to monitor security at the same time if no dedicated security personnel is available.

\section{Application Background and Challenges}
\subsection{Application and Used Data Set}
Our visualization system was developed in collaboration with cyber security experts from the German Research Center for Artificial Intelligence, whose research focuses on detecting attacks on industrial processes from monitoring of OT network sensor readings. This approach is appropriate for distributed OT networks with difficult to maintain or outdated components. Because the infrastructure might not provide basic security features like encryption or authentication, security must be created using the data sources at hand. In our application, these data sources are PLCs monitoring a modern six-stage process of water treatment. The data set containing a real-world underlying process with introduced attacks is provided by the iTrust, Centre for Research in cyber security, Singapore University of Technology and Design \cite{7469060, iTrust.2018}. All attacks are documented, so ground truth comparison of the results is possible.

To provide an overview of our visualization system and its strengths, we consider the development of two components with illustrative results, namely sensor DPIT301 and actuator LIT401. For purpose and location of these components in the considered water treatment process see Section \ref{usage_scenarios} and Figure~\ref{fig:set1}.

\begin{figure*}[t!]
  \centering
  \includegraphics[width=\textwidth]{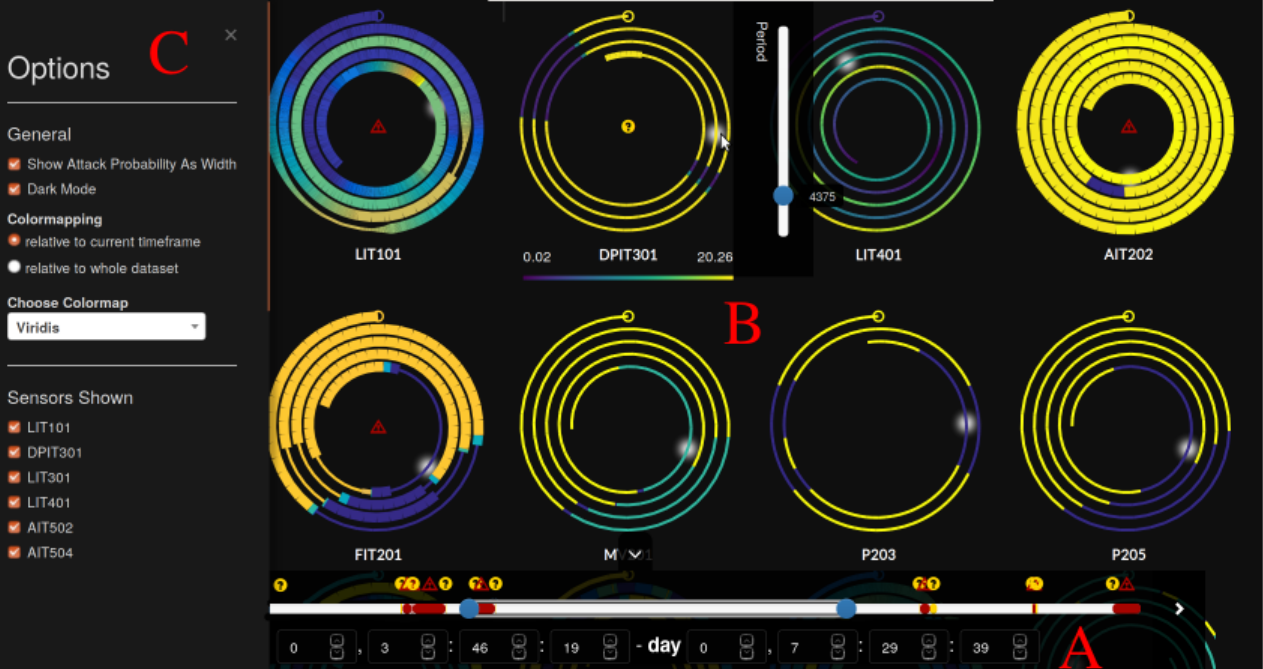}
  \caption{\label{fig:GUI}The visualization system: Time slider [A], spiral chart [B] and options menu [C]. The time slider gives an overview on the complete available data and contained anomalies. In the spiral chart, every sensor is represented by a spiral plot winding from the center to the border. It encodes the measured data in color and the anomaly score as line thickness.}
\end{figure*}

\subsection{Anomaly Detection}
The setting of monitored sensor data in industrial processes differs fundamentally from monitored network activity in IT networks. While in the latter, periodic patterns indicate attacks, in the former the absence of frequent patterns or periodicity indicates anomalies. Thus, three different approaches for anomaly detection based on prediction/comparison of the measurements were implemented and evaluated by Duque Anton et al. \cite{Simon}, namely one-class support vector machines, isolation forests and matrix profiles. While the first two approaches are one-class classifiers that analyze the data on a packet basis, i.e. step by step, matrix profiles are used to perform a time series analysis of a given complete time frame. One-class classifiers require extensive training with a large amount of data. This is time consuming and the data might not be easy to acquire. Even though the detection capabilities of matrix profiles are increased with a larger data base for comparison, they do not require formal training. Furthermore, compared to tuning on hyper-parameters of the used one-class classifiers, matrix profiles require very little tuning. This makes the detection easy to set up and robust to different kinds of data. Based on these facts and the obtained results for the considered data set, matrix profiles turned out to be most suitable for the task of anomaly detection. Hence, only results obtained from this approach are considered here.

Matrix profiles as described by Yeh et al. were developed as an algorithm for motif discovery \cite{Yeh.2016a}. They are based on the comparison of intervals of the analyzed data with the remainder of the data. A distance (e.g. z-normalized distance) between the current sub-series of a given length and all other sub-series of the same length is calculated every time the considered interval is shifted by one. An interactive application of matrix profiles can be found at \cite{mppage}.
% \begin{equation}\label{eq:z_norm_dist}
% \begin{split}
% d(x,y) = \sqrt{\sum_{i=1}^{m}{(\bar{x}_{i} - \bar{y}_{i})}^2} \\
% \bar{x}_{i} = \frac{x_{i} - \mu_{x}}{\sigma_{x}},\quad \bar{y}_{i} = \frac{y_{i} - \mu_{y}}{\sigma_{y}}
% \end{split}
% \end{equation}
% Where $\sigma_{x}$ and $\mu_{x}$ are standard deviation and mean value respectively.
% \begin{equation}\label{eq:mu}
% \begin{split}
% \mu_x = \frac{\sum_{i=1}^{m} x_i}{m}, \quad \mu_y = \frac{\sum_{i=1}^{m} y_i}{m}
% \end{split}
% \end{equation}
% and
% \begin{equation}\label{eq:sigma}
% \begin{split}
% \sigma_{x}^{2} = \frac{\sum_{i=1}^{m} x_{i}^{2}}{m} - \mu_{x}^{2}, \quad \sigma_{y}^{2} = \frac{\sum_{i=1}^{m} y_{i}^{2}}{m} - \mu_{y}^{2}.
% \end{split}
% \end{equation}
% After applying Pearson's Correlation Coefficient as described by Benesty et al.~\cite{benesty2009noise}
% \begin{equation}\label{eq:pearson}
% \begin{split}
% corr(x,y) &= \frac{E((x - \mu_x) (y-\mu_y))}{\sigma_x \sigma_y} \\
% & = \frac{\sum^{m}_{i=1}x_i y_i - m \mu_x \mu_y}{m \sigma_x \sigma_y},
% \end{split}
% \end{equation}
% The Euclidean distance relates as described by Mueen et al.~\cite{Mueen.2010},
% \begin{equation}\label{eq:relation}
% \begin{split}
% d(x,y) = \sqrt{2m(1-\text{corr}(x,y))}
% \end{split}
% \end{equation}
% the resulting metric for distance calculation is then given by
% \begin{equation}\label{eq:working_formular_dist}
% \begin{split}
% d(x, y) = \sqrt{2m\bigg(1-\frac{\sum_{i=1}^{m} x_{i} y_{i} - m \mu_{x} \mu_{y}}{m \sigma_{x} \sigma_{y}}\bigg)}\text{.}
% \end{split}
% \end{equation}
The minimal found distance is then used as measurement for the anomaly of the current interval, the anomaly score. Thus, the algorithm checks if the considered development occurs at other time points as well. To avoid missing attacks that are executed multiple times, a counter for “close” intervals was implemented and taken into account when rating the anomaly probability. If an event is found more infrequent than other events, it is likely to be an anomaly.

The final result of the anomaly detection algorithm is a score for every timestep and every sensor that indicates the probability of an anomaly at this time point. These scores are divided into three categories: 
\begin{enumerate}[label=\Roman*]\setlength\itemsep{-0.3em}
 \item values where an anomaly is unlikely,
 \item values that are extraordinarily high but are not high enough to clearly indicate malicious activity and
 \item values where an anomaly is very likely.
\end{enumerate}
The thresholds used for this categorization can be adapted to the used anomaly detection. If not meaningful in a given context, category II can be omitted from analysis by setting the thresholds accordingly. The thresholds in the presented application have been chosen based on the values obtained from anomaly detection on normal data. Having a small range in which values are classified as category II directly above the range obtained when considering normal data provides a buffer for possible normal values with slightly higher results and thus increases the credibility of category III. While anormal behavior might arise from sources different than malicious activity, it should always be noticed and investigated. Furthermore, the basic idea of using anomaly detection to detect attacks is that attacking the system will lead to anormal values.

\subsection{State-of-the-art analysis} 
Up to now, verification of detected anomalies was performed using basic time series visualizations as in Figure~\ref{fig:anomaly_example}. Five anomalies are detected in the given time frame using matrix profiles. While the matrix profiles detect anomalies with a sufficiently high probability, they do not indicate why an anomaly was detected. Identifying the reason for the anomaly detection is challenging in this setting, as well as the assessment of detected anomalies as true or false positives. For example, moderate changes in the periodical behavior of the data are missed easily. Hence, a more elaborate visualization is needed to support triage analysis, making use of the periodic behavior of sensors and actuators and supporting the understanding of anomaly detection results. 
\begin{figure}[t!]
  \centering
  \includegraphics[height=\spiralheight]{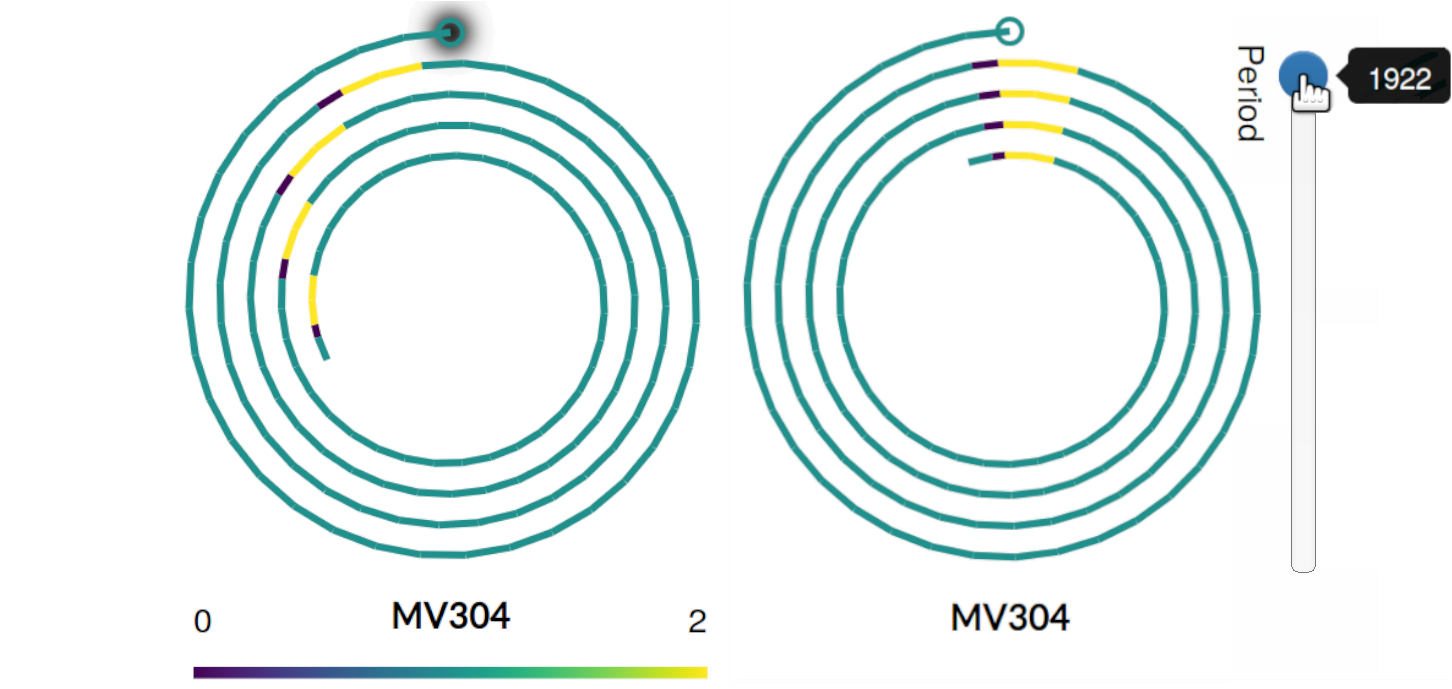}
  \caption{\label{fig:FindPeriod} Interactive adjustment of cycle duration: (left) automatically detected period in seconds, (right) manually adjusted period.}
\end{figure}
\begin{figure}[t]
 \centering
    \includegraphics[height=\spiralheight]{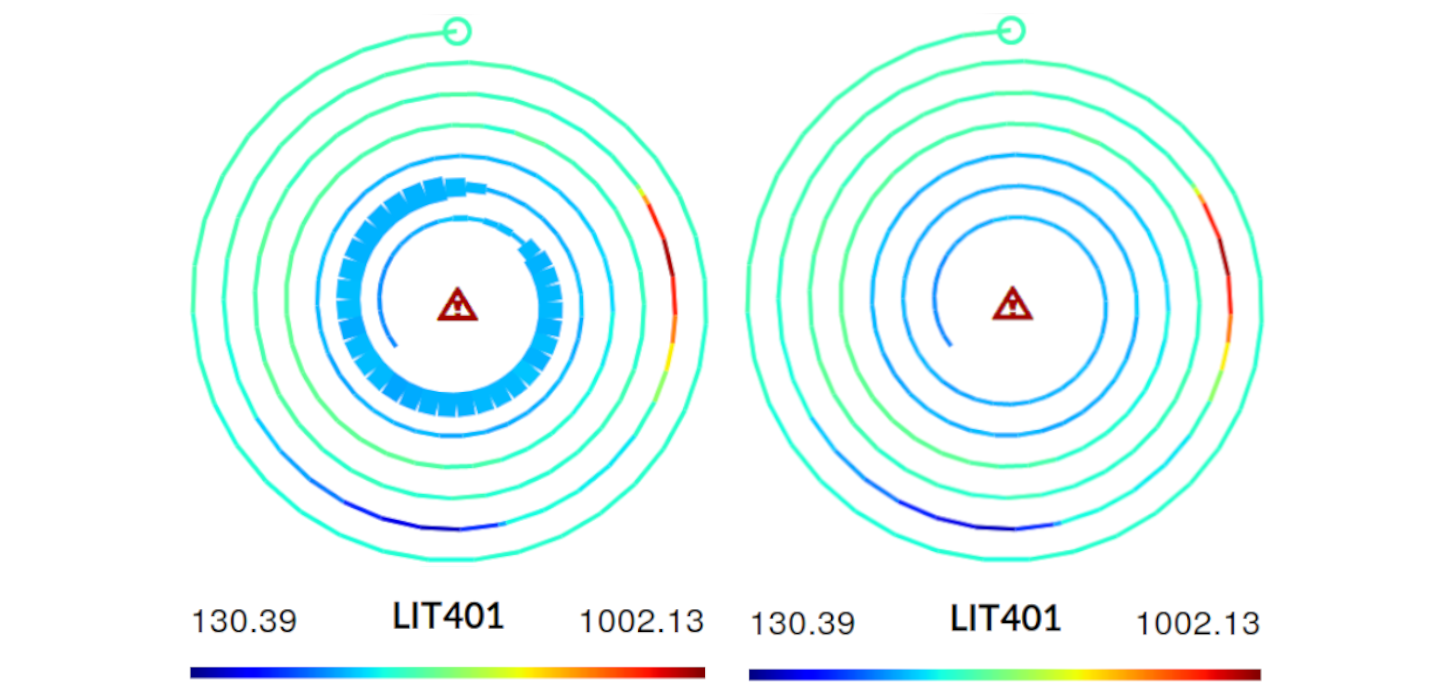}
  \caption{\label{fig:LineSize} Anomaly score visualization: (left) the anomaly score is visualized as line thickness, (right) anomaly score visualization is turned off.}
\end{figure}

\section{System Requirements}\label{requirements}
While anomaly detection is a first step towards an early detection of attacks on the industrial process, its results need to be visualized for verification. 
%Depending on the algorithm used for anomaly detection, live monitoring of the system is possible: Matrix Profiles only allow analysis of a fixed time frame, the other algorithms work on packet basis and can be applied in real time. 
The following requirements for our visualization system have been identified in cooperation with cyber security experts:
\begin{itemize}\setlength\itemsep{-0.3em}
 \item[R1] System monitoring and triage analysis should be supported simultaneously.
 \item[R2] Detected anomalies should be clearly highlighted in the data. 
 \item[R3] Classification of values in category~II as abnormal or normal,
 \item[R4] identification of false positives should be possible using the visualization system.
 \item[R5] The displayed information and the interaction possibilities should allow identification of false negatives. 
 \item[R6] The visualization system should render triage analysis by experts as well as by laymen (in terms of cyber security) possible. 
\end{itemize}
In addition, depending on the algorithm used for anomaly detection, live monitoring of the system should be possible where new values are added to the data set at a given frequency. 

\section{Visualization System}
Our visualization system was designed according to the system requirements given in Section \ref{requirements}. An example application is available at \url{https://priest.cs.uni-kl.de/~lohfink}.

\subsection{User Interface}
The user interface consists of three main parts as shown in Figure~\ref{fig:GUI}. The \textbf{time slider [A]} represents the complete data set and provides an overview and the temporal navigation for the \textbf{spiral chart [B]}. This chart represents the main area of the visualization system. It provides a detailed view on the time interval chosen in the time slider using one spiral plot per selected sensor. The \textbf{options menu [C]} is hidden by default. Here, all visualization options can be chosen and sensors can be included or excluded from visualization. Also, the theme of the whole visualization system can be switched between light and dark mode. 
Figures \ref{fig:FindPeriod} to \ref{fig:attack_mark} illustrate the introduction of this component's features.
Requirements leading to decisions in the design process of single features are named in parentheses.

\subsection{Overview and Detail}
As shown in Figure~\ref{fig:anomaly_example}, the time slider represents the entire provided data and highlights detected intervals of categories II (potential anomaly) and III (anomaly). This is done across all available sensors, providing an overview of the process status (R1, R2). From the complete available data, a time frame can be selected interactively to obtain detailed information in the spiral chart. The maximum size of the selected time frame is set to four hours, covering 14,400 values at one data point per second. This size was chosen based on the experience of the involved security experts. The size of the spiral plots was chosen accordingly. If needed, spiral plot size and the maximum visualized length can be adapted easily. 

The spiral chart contains visualizations of monitored data and anomaly detection for every selected sensor during the selected time frame.  To fully exploit the periodicity of the data and the possibility to detect anomalies visually, we chose spiral plots for this visualization (R1, R3-6) \cite{weber2001visualizing}. Our configuration of spiral plots is the following: 

The time axis is winding from the center of the spiral to the circumference (Figure~\ref{fig:GUI}). That is, the most recent time point is always shown in the largest, outermost circle at the top, highlighted by a small circular marker in corresponding color (R1). This is especially helpful when using the system in live mode, where new data is streamed continuously in the visualization. New data points are added on the largest, outermost circle of the spiral, providing most details for the most recent time interval. Changes in the streamed measurements are accentuated by the change of the marker's color. As the endpoint of the spiral is fixed, widening the visualized time frame results in growing of the spiral towards the center. The earliest visualized time point is represented by the end of the spiral closest to its center. 

The initial cycle length of each spiral is estimated using the zero-crossing method on the corresponding sensor readings. As suggested in \cite{weber2001visualizing} by Weber et al. the cycle length of each spiral can be manipulated interactively, thus taking advantage of the user's pattern recognition abilities to increase the accuracy of the presented period length (see Figure~\ref{fig:FindPeriod}). It also allows an adaption of the cycle length to a change in the data's periodical behavior during analysis (R1, R3-6). Such a change could for example be provoked by operator interventions as explained by Caselli et al. \cite{caselli2015sequence}. An example is given in Section \ref{usage_scenarios}.
Increasing the cycle length, more timesteps are visualized per cycle and the overall spiral becomes shorter with less revolutions. On the other hand, decreasing the cycle length results in fewer timesteps per cycle and thus in a longer overall spiral with more revolutions that ends closer to the center. Since the periodical behaviors of different sensor readings are a priori independent, the period for each spiral plot is treated individually. 

Colors in the spiral represent the different sensor readings either relative to all available measurements obtained by the considered sensor or relative to the currently visualized time frame. The first option is suitable to compare values of the currently selected time frame to the overall values while the second one gives more detail on behavior of the data in the current time frame. Among others, the Parula colormap is available which is suitable for people with color blindness. 

The pre-attentive attribute of line thickness \cite{archambault2015data} is used to provide information on the calculated anomaly score at each time point, i.e. the probability for an anomaly calculated by the employed detector (R2, R6): The thicker a line is, the more likely an anomaly occurred. Values of category~III are of a given maximal thickness, values of category~I of a given minimal thickness and values in between (category~II) interpolate between maximal and minimal thickness. To gain an overview of just the data, the spiral plot can optionally be drawn with a constant line thickness (R1) (see Figure~\ref{fig:LineSize}). 

\begin{figure}[t!]
  \centering
  \includegraphics[height=\spiralheight]{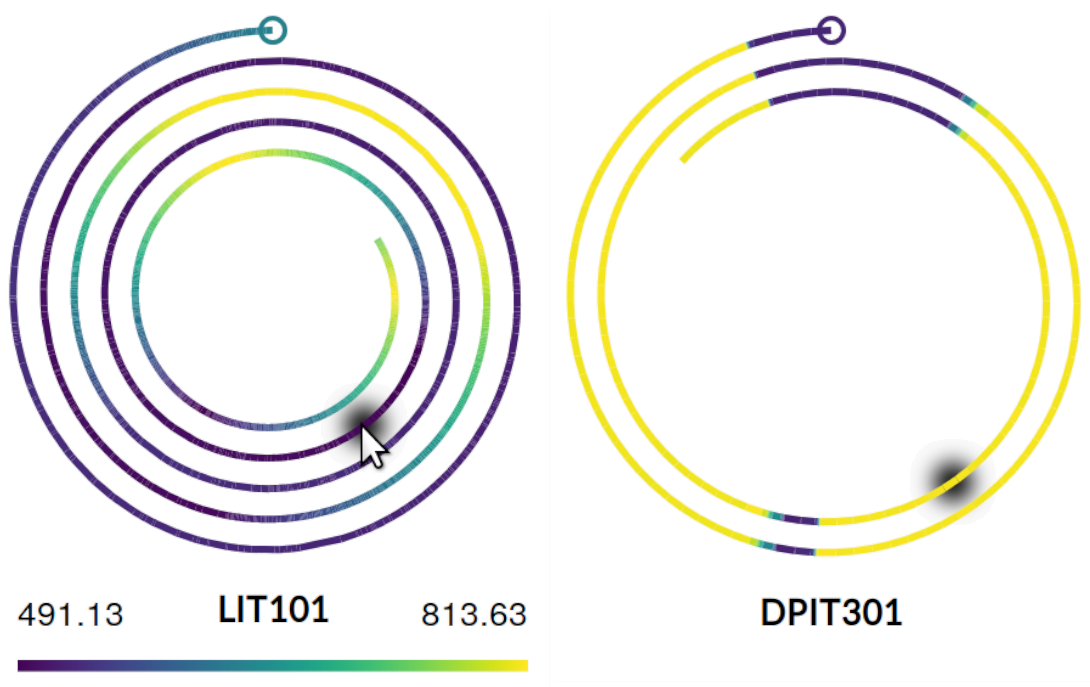}
  \caption{\label{fig:Spotlight} Spotlights: Highlighting the hovered timestep in every spiral simplifies linking of concurrent features.}
\end{figure}

As soon as the cursor hovers over any spiral on screen, the corresponding colormap and value range are shown, and a spotlight follows the cursor to facilitate linking between time points on spirals. This spotlight will always highlight the closest point on the spiral. Simultaneously, spotlights in all other spirals highlight the corresponding time point according to their cycle length in linked views. Due to the fact that every spiral shows the same number of time steps but with different cycle length, outermost and innermost points of all spirals coincide. In between, the speeds of the spotlights differ. Keeping the spotlight behind the actual spiral visualization enables the user to easily compare single data points at each visualized time point (R1, R3-6) (See Figure~\ref{fig:Spotlight}).

\subsection{Anomaly Highlighting}
Detected (potential) anomalies are highlighted in the time slider as well as in the spiral chart (R2). Across the entire visualization system, categories II and III are marked consistently in yellow and red with easily distinguished symbols (question mark and exclamation mark). If both categories II and III are present, the highlighting for the more problematic category~III is chosen.

In the time slider, time frames where any available sensor provided suspicious data are highlighted. The slider bar is colored accordingly in these areas. Centered above the colored area the corresponding symbol is shown. Clicking on this symbol, time frame selection is fitted to the corresponding area. In addition, all sensors where the (potential) anomaly was detected are selected for visualization, ensuring that the users are able to identify all affected sensors. After a time frame including data points from categories II or III was selected, further highlighting is provided in the spiral chart. 

\begin{figure}[t!]
  \centering
  \includegraphics[height=\spiralheight]{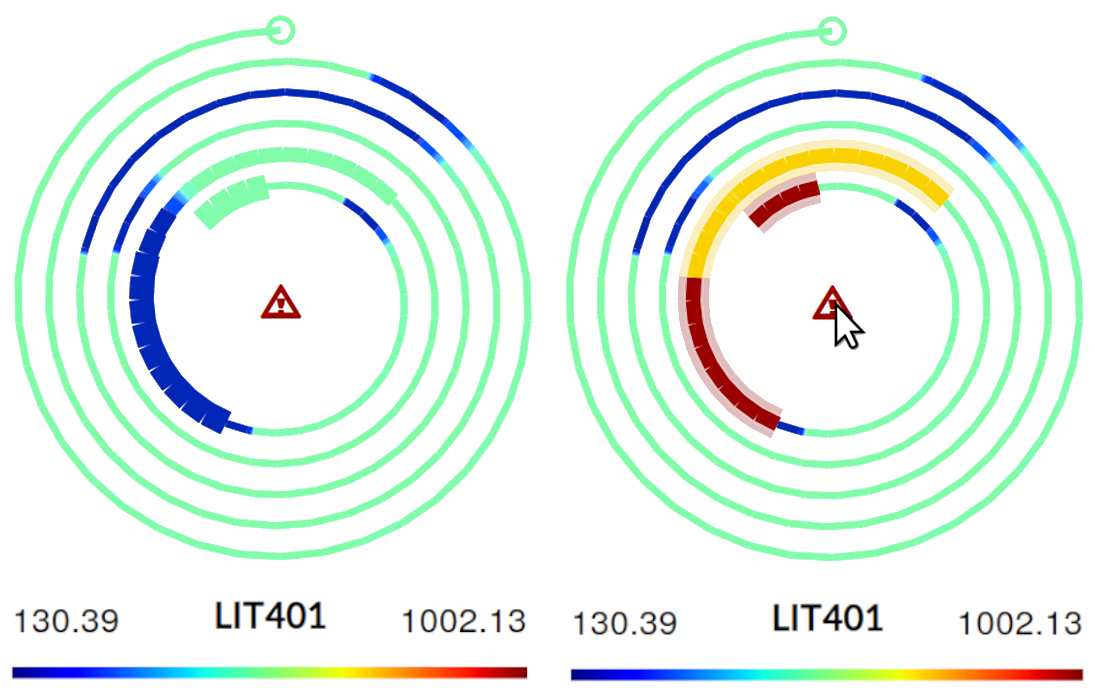}
  \caption{\label{fig:attack_mark} Anomaly highlighting: (left) No highlighting, (right) areas of categories II and III are colored accordingly and line thickness is animated.}
\end{figure}

In this detailed view, to emphasize the occurrence of (potential) anomalies in the data, the corresponding symbol is displayed in the center of spirals that contain data categorized as II or III. See for example sensors LIT101 and DPIT301 in Figure~\ref{fig:GUI}. Especially if the anomaly score visualization is deactivated and the line thickness of the spiral plots is constant, these symbols ensure that no occurring attack is missed. Hovering over the symbol in a spiral's center, data points contained in the spiral plot of categories II and III are highlighted in the corresponding colors. To further draw attention to the highlighted areas and to ensure that the highlighting colors are not mistaken as part of the used colormap, line thickness is animated (see Figure~\ref{fig:attack_mark})~(R3, R4, R6).

\subsection{Interaction}
Multiple interaction opportunities are provided to support monitoring of the process and triage analysis of detected anomalies (R1, R6): 

Navigating the overall data by manipulating the time slider is possible in multiple ways: The borders of the selected time frame can be adjusted by dragging the border markers or clicking on the slider. If the maximum size of the selection is exceeded, the border that is currently not dragged is adjusted to obtain a valid interval. To shift the entire selected time frame while keeping its current size, the arrow buttons next to the slider can be clicked. Precisely known time points (if available) can be entered directly below the slider in the time frame information. Clicking on symbols indicating (potential) anomalies adjusts the time frame to include all corresponding time points and selects all affected sensors for visualization. 

After the time frame is fixed, the plots on the spiral chart can be adjusted individually. Clicking on a spiral, a slider to interactively manipulate the spirals cycle length is shown. Options applying to all shown spiral plots are available in the options menu. The colormap and its borders can be changed and visualizing results from anomaly detection as line width can be toggled. To be able to focus on individual sensor readings, uninteresting sensors can be excluded from the visualization.  


\subsection{Implementation and Scalability}
The visualization system was implemented using web technologies (HTML, CSS, and JavaScript) in the interest of portability and maintainability. The controls are chosen to permit usage of the visualization system on touch screens. 
Significant speedups compared to a na\"ive implementation were gained by rendering the spirals using a given threshold $\epsilon$ and a maximal line length $k$. A line is started at the first drawn timestep and traced step by step with step size defined by the spiral's cycle length. As long as the first encountered data point of the line does not differ more than $\epsilon$ from the following data points, a single line with a single color is drawn. The line either ends when a higher difference occurs or the maximal length $k$ is reached. Hence, there are less lines drawn than timesteps are contained in the spiral, leading to a speedup in rendering time. 
This precludes a highlighting of single timesteps in the spiral, therefore the spotlight calculation does not rely on drawn elements but is purely analytical. 
Spirals were implemented using the HTML SVG element, re-calculation and re-drawing were minimized. In our tests, 51 different sensors were shown at a time and smooth real time-interaction was possible. This holds also true when settings are changed.\par
Showing significantly more than the 51 sensors would result in longer loading times, and scrolling a long page full of sensors would not be optimal. Having different sub-pages with a meaningful division of the sensors would facilitate the navigation, provide a better overview and allow the system to asynchronously load additional sensor data. Thus interactivity on single pages is assured. To mark potential threats on different sub-pages, the established icons and colors could be used. This is part of our future work.

\section{Analysis Strategies and Usage Scenario} \label{usage_scenarios}
\begin{figure}[t!]
  \centering
  \includegraphics[width=\columnwidth]{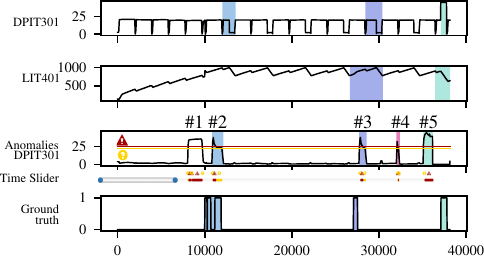}
  \caption{\label{fig:anomalydetectionSet1}Anomaly detection results (extended Figure~\ref{fig:anomaly_example}): Ground truth is provided and detected anomalies are highlighted.}
\end{figure}

\begin{figure}[t!]
  \centering
  \includegraphics[width=\columnwidth]{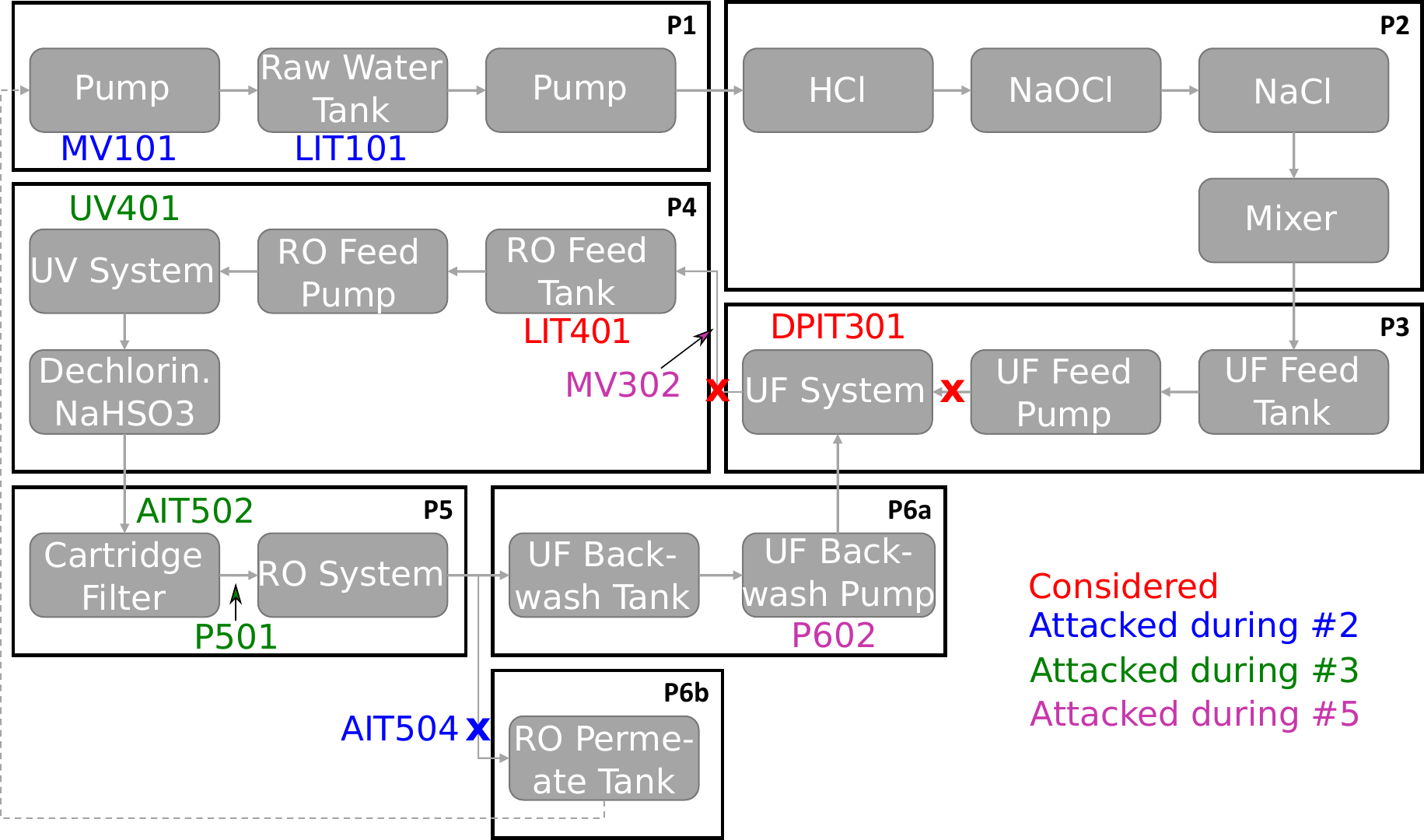}
  \caption{\label{fig:set1} Sub-processes of the used data set and their interplay: Attacked components and the sensor and actuator considered in the usage scenario are highlighted.}
\end{figure}

To evaluate results from the anomaly detection in Figure~\ref{fig:anomalydetectionSet1}, our visualization system was employed. The considered data set contains sensor readings from a modern six-stage process of water treatment. 
The monitored sub-processes are:\\
\textbf{P1}: Raw water storage,\\
\textbf{P2}: Pre-treatment,\\
\textbf{P3}: Membrane Ultra Filtration (UF),\\
\textbf{P4}: Dechlorination by Ultraviolet (UV) lamps,\\
\textbf{P5}: Reverse Osmosis (RO),\\
\textbf{P6}: Disposal\\
% \begin{itemize}
% \item[P1] Raw water storage
% \item[P2] Pre-treatment
% \item[P3] Membrane Ultra Filtration (UF)
% \item[P4] Dechlorination by Ultraviolet (UV) lamps
% \item[P5] Reverse Osmosis (RO)
% \item[P6] Disposal
% \end{itemize}
Connections between the sub-processes are shown in Figure~\ref{fig:set1}.

The digital part of the system consists of a layered OT-network, PLCs, human machine interfaces and a \textit{Supervisory Control And Data Acquisition} (SCADA) workstation. PLCs are industrial digital computers that are tailored to controlling manufacturing processes and can be connected to sensors and actuators. In total, 51 sensors and actuators are controlled by the PLCs. A detailed list including a description can be found in \cite{Goh.2016}. Here, also the attacks contained in the data set and their occurrences are described.

Sensors DPIT301, the differential pressure indicating transmitter controlling the back-wash process in stage P3, and the actuator LIT401 reporting the water tank level of the tank in stage P4 are considered in this usage scenario. See Figure~\ref{fig:set1} for the location of these components in the overall process. The analyzed data consists of 10,000 samples of normal behavior as a ``base line'' to compare occurring patterns to, and 28,000 samples containing five attacks. Five anomalies have been detected in the readings of sensor DPIT301. For sensor LIT401, just one anomaly, coinciding with the first anomaly in DPIT301, was detected (not shown here). This first detection is an artifact at the edge between normal data and data containing attacks. Furthermore, the 4th detection in the readings of DPIT301 is a false positive. In Figure~\ref{fig:anomalydetectionSet1} the detected anomalies and corresponding areas in the sensor readings are highlighted. In addition, the ground truth is given, where value 1 stands for an attack. 
For every anomaly, a different analysis strategy that is supported by our visualization tool is applied. The five exemplary situations are: %\\
%\textbf{Period disruption}: A sudden change in the period of the measurements occurs.\\
%\textbf{Abnormal occurrence of values}: Values that are in the normal range occur at an abnormal time point or for an abnormal length.\\
%\textbf{Phase shift}: The period of the measurement changes phase shift.\\
%\textbf{False positive}\\
%\textbf{Abnormal values}: Values that are not within the normal range occur.\\
 \begin{itemize}
 \setlength\itemsep{0.0em}
  \item Period disruption: A sudden change in the period of the measurements occurs.
  \item Abnormal occurrence of values: Values that are in the normal range occur at an abnormal time point or for an abnormal length.
  \item Phase shift: The period of the measurement changes phase shift. 
  \item False positive
  \item Abnormal values: Values outside the normal range occur.
 \end{itemize}

\begin{figure}[ht!]
  \centering
    \includegraphics[height=\spiralheight]{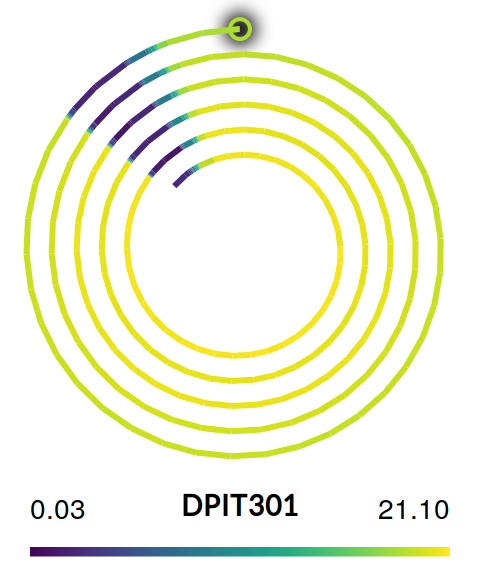} 
    \caption{\label{fig:1_previous_period}The period of DPIT301 before the first anomaly.}
  \end{figure}

\begin{figure}[ht!]
  \centering
    \includegraphics[height=\spiralheight+0.2cm]{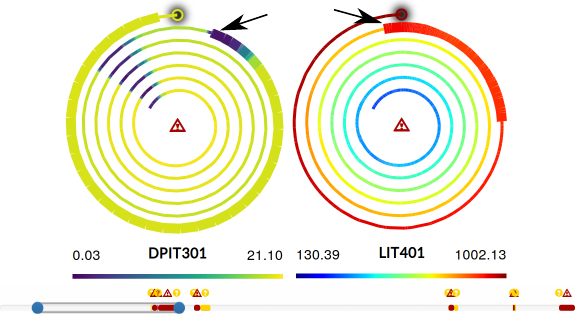} 
    \caption{\label{fig:1_anomaly}Anomaly \#1: (left) Disrupted period, (right) abrupt increase.}
  \end{figure}

The peaks were evaluated one by one as follows:\\
\\
\textbf{Detected anomaly \#1, period disruption:}
This detection is caused by the edge between normal data and data containing attacks. The combination of two data sets caused a period disruption and a sudden change in values.\\
In our visualization system both effects are visible: 
The period of DPIT301 in the normal data previous to any attacks was perfectly identified by the period estimation (Figure~\ref{fig:1_previous_period}).
Widening the considered time interval to include the detected anomaly provides insight in the reasons for the detection (Figure~\ref{fig:1_anomaly}): Clearly, the period of DPIT301 is disrupted in the area that was detected as abnormal (left). In addition, the small, abrupt increase in the values of LIT401 is visible, especially if the jet colormap is chosen (right).
\\
\textbf{Detected anomaly \#2, abnormal occurrence of values:} 
Three short attacks in rapid succession lead to this abnormal behavior. Two attacks were executed on permeate conductivity analyser AIT504 measuring the NaCl level in sub-process 5 at the reverse osmosis system. The expected outcome was missed and the attacks did not lead to much change in the data. The third attack was on the motorized valve MV101 that controls the water flow to the raw water tank at the beginning of the process and level transmitter LIT101 controlling the raw water tank level. The valve is kept on while the transmitter is kept on a constant level, avoiding a shutdown of the valve. This leads to an overflow of the raw water tank and propagates further through the system, causing lagging.

\begin{figure}[ht!]
  \centering
    \includegraphics[height=\spiralheight]{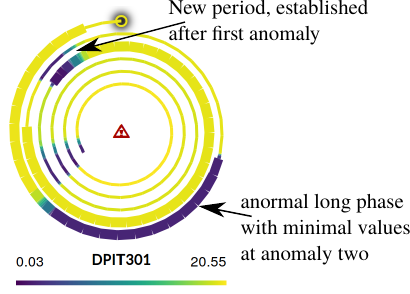} 
    \caption{\label{fig:2_period}Anomaly \#2: An extraordinarily long phase with minimal values for DPIT301.}
  \end{figure}

\begin{figure}[ht!]
  \centering
    \includegraphics[height=\spiralheight]{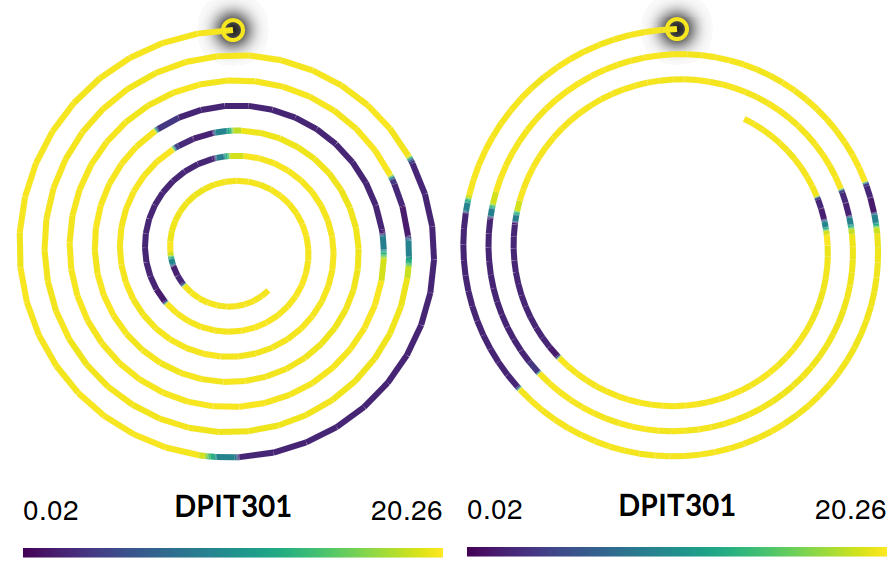} 
    \caption{\label{fig:3_adapt_period}Cycle length adjustment: (left) Automatically detected period length, (right) after manual adjustment.}
  \end{figure}

\begin{figure}[ht!]
   \centering
    \includegraphics[height=\spiralheight]{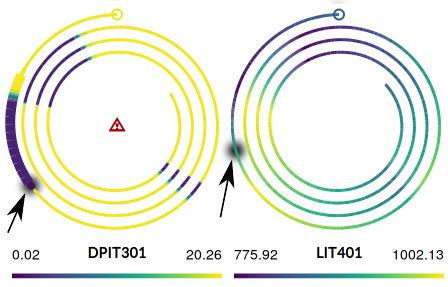} 
    \caption{\label{fig:3_shifts}Anomaly \#3: Period shift}
  \end{figure}

\begin{figure}[ht!]
  \centering
    \includegraphics[height=\spiralheight]{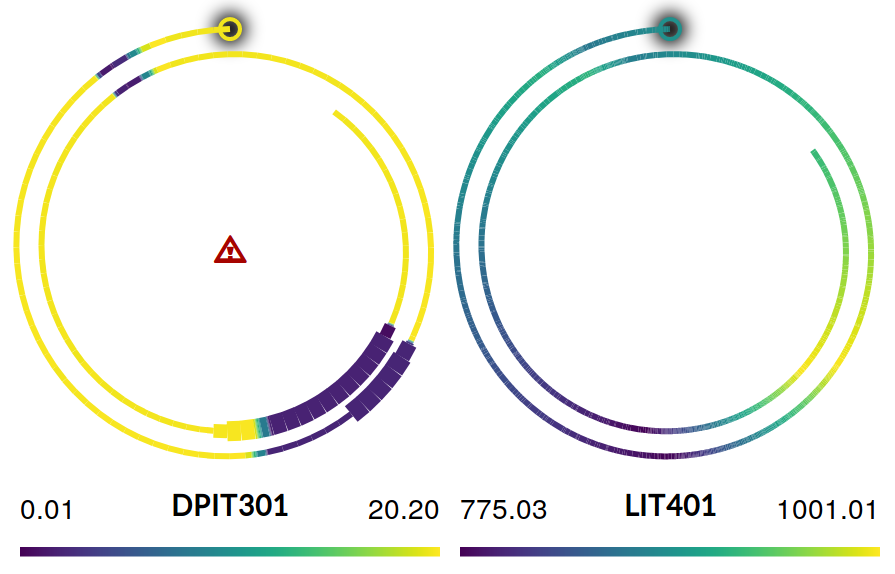} 
    \caption{\label{fig:false_attack} Anomaly \#4 is a false positive. No anormal behavior can be spotted visually.}
  \end{figure}

This lagging is clearly visible using our system: 
After anomaly \#1, the readings of DPIT301 re-start their periodical behavior with a phase shift. After one cycle, the next anomaly is detected, resulting in an extraordinarily long phase with low values. Although the new period is not yet well established, this phase clearly stands out from the data and is obviously the reason for the detected anomaly (Figure~\ref{fig:2_period}). Hence the visualization leads to the correct classification of the detected anomaly as an attack. 
\\
\\
\textbf{Detected anomaly \#3, phase shift:} 
Here, the abnormal behavior was triggered by an attack on the dechlorinator UV401 in sub-process 4 used to remove the chlorine from the water. It was stopped and the value of oxidation-reduction potential analyser AIT502 monitoring the NaOCl level in the reverse osmosis feed was fixed to prevent an alert. In addition, the pump P501 pumping dechlorinated water to the reverse osmosis system was kept on. During the attack it was not possible to force P501 to stay on; so a possible damage was avoided and the chlorine loaded water was rejected at sub-process 6. This leads to higher input in the ultrafiltration process which is visible in the readings of DPIT301 and LIT401 using our system after an adaption of the period:\\
After anomaly \#2, the period of DPIT301 from the second data set containing the attacks is fully established. The estimated cycle length is no longer valid and the periodicity is not easy to spot. The cycle length of the visualization can be easily adapted so that a further analysis of the period is possible (Figure~\ref{fig:3_adapt_period}).\\
\\
At detected anomaly \#3, periods of DPIT301 and LIT401 change the phase shift. Although in LIT401 no anomaly was detected, this shift is clearly visible in our visualization system after adapting the cycle length (Figure~\ref{fig:3_shifts}). This identifies the detected anomaly as an attack (which is confirmed by the ground truth). The phase shift of LIT401 is also difficult to spot in Figure~\ref{fig:anomalydetectionSet1}, leading us to the impression that our visualization system is indeed superior to na\"ive time series visualizations.\\
  \\
\\
%\vfill\null
%\columnbreak
\textbf{Detected anomaly \#4, false positive:} 
Around detected anomaly \#4, no changes in values or periods are visible. Hence, an attack at this time point is doubtable. In fact, this detected anomaly is a false positive that can be identified fairly easy using our visualization system (Figure~\ref{fig:false_attack}). 
\\
\textbf{Detected anomaly \#5, abnormal values:} 
The triggering attack for this anomaly detection involved keeping the backwash pump P602 in sub-process 6 closed, setting the value of DPIT301 to a high value and keeping the motorized valve MV302 that controls the flow from the ultrafiltration process to the de-chlorination unit open. This leads to a system freeze since no water from the backwash pump is available while water transport from sub-process 3 to 4 is kept active. The reverse osmosis feed tank runs dry.

\begin{figure}[ht!]
  \centering
    \includegraphics[height=\spiralheight]{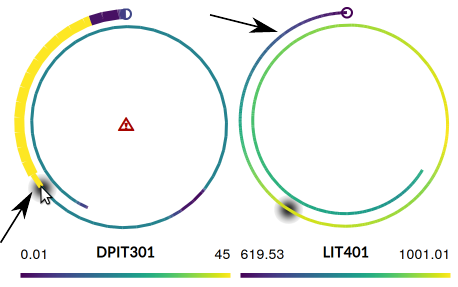} 
    \caption{\label{5_outlier} Anomaly \#5: (left) sudden increase in DPIT301, (right) rapid decrease in LIT401.}
  \end{figure}

Again, this attack is clearly visible in our system. Even tracing the origin of the attack is possible: 
At the time frame of detected anomaly \#5, the values of both sensors change tremendously and fast, indicating an attack. Choosing the colormap relative to the current time frame, this is easy to see for both sensors, making it easy to correctly classify this anomaly as an attack (Figure~\ref{5_outlier}).
  
The direct attack on DPIT301 and also the drop in values in LIT401 caused by the dry running tank are clearly visible.
In addition, the succession of visible effects that can be traced using the spotlights indicates that the attack first affected DPIT301 and propagated to LIT401 afterwards. This provides a hint on the origin of the attack. At the beginning of the manipulation of values in DPIT301, the anomaly detection provides values from category II, as the value remains unusually high, category III is detected. This increase in values (and thus line thickness) is shown clearly. Hovering the alert icon at the spiral center, the segment of category II will be highlighted in yellow, the segment of category III in red. 

Overall, our visualization system proved useful in supporting cyber security experts in their triage analysis tasks. Feedback of the experts is given in Section \ref{experteval}.

\section{Evaluation} 
\subsection{Qualitative Evaluation}
Our visualization system was evaluated in cooperation with the HCI group at the Technische Universit\"at Kaiserslautern. To assess requirement R6, the ability of laymen (in terms of cyber security) to verify results of the anomaly detection, the evaluation was performed with 15 subjects, where 11 have a technical background (IT/electrical engineering) and none have experience in cyber security. This is also in consent with the findings of Staheli et al. in \cite{Staheli:2014:VEC:2671491.2671492}. After a short introduction to the system, the users performed several tasks and filled the questionnaire on effectiveness along the way. Afterwards, they were asked to fill a questionnaire on usability of the system. 

\subsubsection{Effectiveness Results}
The requirements defined in Section \ref{requirements} were assessed based on a questionnaire with multiple tasks to perform on the secure water treatment data set. One false positive alert was added to the anomaly detection data and one alert was deleted to create a false negative. Areas of category~II belonging to categories I and III were already present in the data. \\
Participants were asked to: 
\begin{itemize} \setlength\itemsep{-0.3em}
 \item [Q1]identify a current threat,
 \item [Q2]determine the periods of some given measurements,
 \item [Q3]determine the sensors that are affected by an attack,
 \item [Q4]identify true and false positives,
 \item [Q5]classify areas of category~II as category~I or category~III and
 \item [Q6]spot suspicious measurements in an area detected as normal.
\end{itemize}
Q1 and Q2 assess R1, the ability to monitor the system and simultaneously perform triage analysis. Requirement R2, that requires detected anomalies to be clearly highlighted in the data is assessed by Q1 and Q3. The classification of values in category~II (R3) is assessed in Q5, once for each category. The ability to detect false positives with our system (R4) is tested in Q4 once with a true positive, and once with a false positive. R5, support of the system to detect false negatives, is assessed in Q6.

The results for each question are presented in Table \ref{table:qualitative_results}.
\begin{table}[]
\begin{center}
\begin{tabularx}{\columnwidth}{|l|X|l|l|}
\hline
\multicolumn{2}{|c|}{Question}                                                                                                  & \checked & SD \\ \hline
Q1a & Identify a current anomaly                                                                                                 & 100\%   & 0                                                             \\ \hline
Q1b & Determine first time of occurrence of a thread                                  & 79\%    & 0.43                                                          \\ \hline
Q2a & Determine the period of a sensor with a single peak per period              & 64\%    & 0.5                                                          \\ \hline
Q2b & Determine the period of a sensor with multiple peaks per period             & 50\%    & 0.5                                                          \\ \hline
Q3  & Determine the sensors that are affected by an anomaly                        & 93\%   & 0.27                                                            \\ \hline
Q4a & Identify a true positive                                                                                                  & 93\%    & 0.27                                                          \\ \hline
Q4b & Identify a false positive                                                                                                 & 85\%    & 0.38                                                           \\ \hline
Q5a & Classify an area of category~II as attack                                      & 100\%   & 0                                                             \\ \hline
Q5b & Classify an area of category~II as normal                                      & 93\%    & 0.27                                                          \\ \hline
Q6a & Spot suspicious measurements in an area that was detected as normal & 100\%   & 0                                                             \\ \hline
Q6b & Determine the sensor with abnormal behavior                                     & 93\%    & 0.27                                                          \\ \hline
\end{tabularx}
\end{center}
\caption{\label{table:qualitative_results}Amount of successful participants and standard deviation.}
\end{table}
All participants were able to identify a current potential thread in the system (Q1a) and 79\% were able to determine the first time point a thread occurs in the overall data correctly (Q1b). The period of sensor data at a given time interval was determined correctly by 64\% in the case of a period with a single peak (Q2a) and by 50\% in the case of a period with multiple peaks per period (Q2b). While these numbers are comparably low, 93\% of the participants were able to detect some kind of pattern for a chosen spiral in the spiral plot. Hence, the advantage of spiral plot visualization concerning the detection of disrupted or shifted periods is still present, even though the correct period was not found. These results and the fact that 93\% of the participants were able to correctly determine the sensors that were affected by a detected anomaly (Q3) lead us to towards opinion that requirements R1 and R2 were implemented successfully in our system. Considering the classification of values in category~II in categories I and III, 100\% of all participants were able to determine an actual attack (Q5a). The decision that an area of category~II is actually normal (Q5b) was made correctly by 93\%. With these results, we consider requirement R3 as met. 93\% of all participants were able to identify a detected anomaly as a true positive (Q4a) and the detection of false positives (Q4b) was performed with 85\% success rate by the participants. Hence, also requirement R4 is met by our system. The supporting character of the system in performing security related tasks is further assessed in the usability study in Section \ref{usability}. Requirement R6 (support of the system in finding false negatives) was assessed by having the participants navigate to an area marked as normal by the system that actually contained an attack. 100\% of all participants were able to identify this attack (Q6a) and 93\% were able to determine the sensor with abnormal behavior correctly (Q6b). Since this question was last in our questionnaire, we see this perfect result as a hint to an existing training effect, that was already present in a 10 minute survey. Also we consider requirement R5 as met. Requirement R6, rendering verification by laymen (in terms of cyber security) possible, was obviously met since all participants were able to understand their tasks and perform most of them correctly.
Overall, the participants navigated confidently through the data in our visualization system and were able to make use of the provided features to perform the tasks even after a very short walk through. The visualization was described as pleasant, the dark visualization mode was a clear favorite. 

\begin{figure}[t!]
  \centering
  \includegraphics[width=\columnwidth]{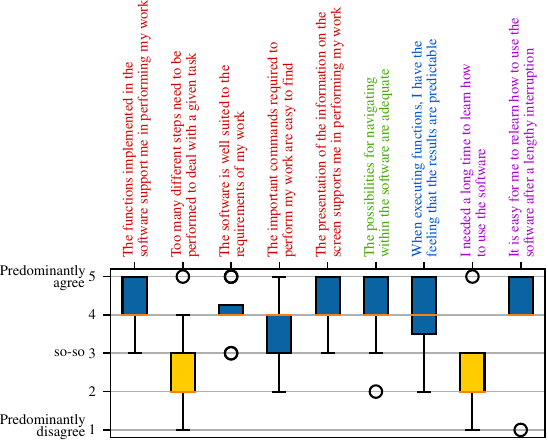}
  \caption{\label{fig:usability_results} Results (quartiles, median, data range and outliers) per question in the usability questionnaire. For questions with blue boxplot, high values are a good result, questions with yellow boxplot require low scores for a positive result.}
\end{figure}

\subsubsection{Usability Results}\label{usability}
Participants of the experiment were also questioned for an assessment on the usability of the visualization system. 
The usability questionnaire according to ISO 9241/10 contained questions about:
\begin{itemize} \setlength\itemsep{-0.3em}
 \item suitability for the task (\textcolor[RGB]{212,0,0}{red}),
 \item controllability (\textcolor[RGB]{68,170,0}{green}),
 \item conformity with user expectations (\textcolor[RGB]{0,85,212}{blue}) and
 \item suitability for learning (\textcolor[RGB]{170,0,212}{purple}).
\end{itemize}
For each question, the participants were asked to respond on a five-level Likert scale from 1 (Predominantly Disagree) to 5 (Predominantly Agree).
The aggregated results for the dimensions mentioned above are presented in Figure~\ref{fig:usability_results}.
Suitability of the visualization system for the performed tasks was evaluated as very good by the participants. In their opinion, the well-suited and supported commands are relatively easy to find and not too many steps are necessary to achieve results. Controllability was also rated as very good as well as conformity with user expectations. Several users suggested to improve the navigation in the time slider by providing a possibility to enter timestamps directly. This feature was added after the evaluation. The previous observation that it is easy to learn how to control the visualization system is confirmed by the participants impressions.

The effectiveness and usability assessment results shown are positive. We interpret these results to indicate that the metaphors of our visualization system are well chosen and that triage analysis in the industrial context can be effectively supported by it.

\subsection{Expert Evaluation}\label{experteval} 
To evaluate usability and usefulness of our visualization system for cyber security experts, an expert evaluation was performed. In the context of a presentation and application of our system, an expert was interviewed. Considering usability and visual presentation of the application, the expert's opinion was very positive. In his judgement, identification of correctly and incorrectly detected anomalies was easy to accomplish using our system. Especially the different employed highlighting techniques for anomalies and potential anomalies were convincing. Also the overall data representation provided a good overview and the maximal length of the visualized time frame is well chosen. The expert stated that our system could be beneficial not only in an industrial cyber security context, but also for process-level monitoring in industrial applications when used to augment existing monitoring systems. 

As an enhancement of our system, the expert suggested the possibility to not only visualize continuous time frames but to allow comparison of discrete time frames. An application example would be periodic behavior of a process that is not on an hourly scale but develops within several days. Furthermore, recurring attacks performed by staff or external employees are conceivable. These would occur only during certain times of the day, or even at certain days of the week. Hence, providing the possibility to show spiral plots not only per sensor or actuator but also per selected time frame is part of our future work. 

\section{Discussion}
In the example usage scenario we found that triage analysis using our tool was effective, comfortable and superior to na\"ive time series visualizations. Each correct anomaly detection could be explained by different, clearly visualized features in the data. Missing these features for the false positive easily leads to rejecting the detected anomaly at hand. Being able to switch the colormap boundaries from ``relative to the whole data set'' to ``relative to the current time frame'' turned out to be helpful when considering either long-term development in the data or local behavior, respectively. Clear accentuation of anomalies using icons, line thickness and highlighting with determined colors leads to an easy to understand overview of the data and potential anomalies. Also, we found that periodicity could be monitored effectively and enabling the user to change the period of the spiral plots not only increases accuracy of the shown period but also allows the user to react to changes in the period during the process. 

The evaluation with laymen w.r.t. cyber security clearly indicates that basic triage analysis tasks can indeed be performed by them by use of our visualization system. Also, monitoring the readings of available sensors in order to gain an overview of the current process was possible. We are convinced that staff that is trained to work with the according machine would perform even better in monitoring and detecting anomalies in the behavior and could benefit considerably from our visualization tool.

Hence, we believe our visualization system and anomaly detection would be a useful tool to
make available not only for incident response or cyber threat hunting teams working in the Operational Technology environment, but also for industrial companies without dedicated cyber security staff. In general, industrial control systems providing anomaly scores on periodical data could profit from our visualization approach.

While all requirements of the given setting were met successfully, requirements to specifically support comparative analysis are not yet incorporated. For example, comparing anomalies from two separate time windows is not possible. In addition, analysis of inter-sensor relations are currently only supported by the spotlights feature. Widening our system by these abilities is an interesting task for future work. Currently, it is up to the user to choose meaningful timeframes from the data witch might not result in optimal choices. Supporting the user by suggesting possible choices could further facilitate the handling of the system. Up to now, the system was designed as a stand-alone solution that could be available directly at an operative unit. Thus, there is no possibility to incorporate or share gained knowledge on previous anomalies and attacks into the system. To provide these possibilities and to enable the user to share results with other analysts is a big part of our future work that will greatly benefit from the browser-based implementation of the system. In the current implementation, the visualization of the sensors is not connected to the physical location of the sensors in the machines due to space-reasons. Splitting the sensors in meaningful groups (f.ex. one per sub-process) could solve this problem and allow the inclusion of references to the real sensor locations in the visualization. In addition, this will improve the scalability of the system.

\section{Conclusion and Future Work}
In this paper we presented a novel visualization system for triage analysis and monitoring of Operational Technology networks, based on a novel anomaly detection in these networks. The visualization system exploits typical patterns that are often inherent in sensor and actuator readings from industrial processes to the fullest to enable cyber security experts and laymen to perform triage analysis and monitoring of the system simultaneously. The main characteristic of our visualization system is manipulable spiral graphs that combine the visualization of sensor readings in their coloring with the results from anomaly detection in their line thickness. Anomalies are highlighted using further pre-attentive properties like form, movement and dedicated colors. 
We presented an example usage scenario, demonstrating the usefulness and effectiveness of our system in triage analysis and monitoring.  
Our system's effectiveness was evaluated in a small user study by cyber security laymen and an expert evaluation was performed. The results were very promising and were discussed in detail.

Further development of our system will involve widening of its scope of application and extension of tools to support comparison of two separate anomalies from two separate time windows and inter-sensor relations. One option to implement this would be by superimposing spiral plots. Proposing meaningful timeframes based on the results from anomaly detection could further improve the usability of the system and will be considered as well. Providing possibilities to incorporate gained knowledge in the system and to cooperate with other users will be considered. Based on the browser implementation, this could span from annotating and marking interesting timeframes to storing data snippets of interest for comparison and adapting thresholds and anomaly detection of the system dynamically. Finally, a referenced visualization of physical sensor locations in the machines to support orientation and consequently the meaningful division of the sensors in groups is part of our future work. Moreover, embedding operational constraints in the visualization could be possible, e.g. by putting sensor visualizations into context of an annotated layout plan containing these constraints.

\acknowledgments{
The authors wish to thank No\"el Str\"ohmer-Lohfink for support in front end development. This paper was funded by the Deutsche Forschungsgemeinschaft (DFG, German Research Foundation) – 252408385 – IRTG 2057.\\
This work has been supported by the Federal Ministry of Education and Research of the Federal Republic of Germany (Foerderkennzeichen 16KIS0932, IUNO Insec). The authors alone are responsible for the content of the paper.}
\vspace{1cm}
\bibliographystyle{abbrv-doi}

\bibliography{SecurityInProcess}
\end{document}